\documentclass[pra,twocolumn,amsmath,amssymb,superscriptaddress,showpacs]{revtex4}
\usepackage{graphics}
\usepackage{epsfig}
\usepackage{bbm}
\usepackage[latin1]{inputenc} 

\newcommand{\be}{\begin{equation}}
\newcommand{\ee}{\end{equation}}
\newcommand{\eea}{\end{eqnarray}}
\newcommand{\bea}{\begin{eqnarray}}

\newcommand{\mean}[1]{\ensuremath{\langle{#1}\rangle}}

\newcommand{\eins}{\ensuremath{\mathbbm 1}}
\newcommand{\qed}{\ensuremath{\hfill \Box}}

\newcommand{\ketbra}[1]{\ensuremath{| #1 \rangle \langle #1 |}}

\newcommand{\ket}[1]{\ensuremath{|#1\rangle}}

\newcommand{\kommentar}[1]{}
\renewcommand{\vr}{\ensuremath{\varrho}}

\begin{document}
\title{
Multipartite entanglement in spin chains
}
\date{\today}
\begin{abstract}
We investigate the presence of multipartite entanglement in 
macroscopic spin chains. We discuss the Heisenberg and the 
XY model and derive bounds on the internal energy for 
systems without multipartite entanglement. Based on this we 
show that in thermal equilibrium the above mentioned spin systems
contain genuine multipartite entanglement, even at 
finite modest temperatures.
\end{abstract}

\author{Otfried G\"uhne}

\affiliation{Institut f\"ur Quantenoptik und Quanteninformation,
\"Osterreichische Akademie der Wissenschaften,
A-6020 Innsbruck, Austria,}

\author{G\'eza T\'oth}

\affiliation{Max-Planck-Institut f\"ur
Quantenoptik, Hans-Kopfermann-Stra{\ss}e 1, D-85748 Garching, Germany,}

\author{Hans J. Briegel}

\affiliation{Institut f\"ur Quantenoptik und Quanteninformation,
\"Osterreichische Akademie der Wissenschaften,
A-6020 Innsbruck, Austria,}

\affiliation{Institut f\"ur Theoretische Physik,
Universit\"at Innsbruck, Technikerstra{\ss}e 25,
A-6020 Innsbruck, Austria}



\maketitle

\section{Introduction}

The study of entanglement in condensed matter systems has become
a major line of research in quantum information science.
In these studies, different aspects of entanglement
have been investigated. First, by investigating
the entanglement in the reduced state of two qubits 
in a spin chain, it has been shown that entanglement 
theory can help to understand the physical properties 
of systems undergoing quantum phase transitions \cite{oster}.
A slightly different approach studies the entanglement 
between two qubits which can be generated by 
local operations on the remaining qubits, the so-called 
localizable entanglement \cite{localizable}.
A third field of study investigates the entanglement 
between a block and the remaining qubits in the ground 
state. The entanglement can be quantified by the entropy
of the reduced state and then the question arises, whether
the entanglement scales like the surface of the block 
\cite{arealaws}. Furthermore, the insights from entanglement 
theory made it possible to understand the limits of known simulation
techniques (like the density matrix renormalization group) 
in condensed matter physics as well as the design of new techniques
which are superior to the ones known before \cite{vidal}. Finally,
it has been shown that macroscopic properties of solids can be
related to entanglement properties of microscopic degrees of
freedom, allowing for a detection of entanglement in macroscopic
objects by observing macroscopic observables 
\cite{buttiker, toth1, remark5}.

In all of the research lines specified above, entanglement
was studied as a {\it bipartite} phenomenon;
the investigation of {\it multipartite} entanglement in spin chains
has only a limited literature \cite{multi1,multi2}.
In entanglement theory, however,  multipartite entanglement has
been intensively studied and it is known that quantum correlations
in the multipartite setting have a much richer structure than in
the bipartite setting \cite{duer, acin}.
The main results known
are for three or four particles, results for large ensembles of
particles are still rare. Nevertheless, studying multipartite
entanglement in large systems can already benefit from the results
for small systems in two respects. First, on a more local scale,
one can pick up sub-ensembles of three or four particles and
investigate the multipartite entanglement of the corresponding
reduced state. Second, on a more global scale, one can ask what
types of multipartite entanglement are necessary to form a given
state of the total system.

In this paper we investigate multipartite entanglement in
macroscopic spin systems. 
The tool we use is similar to the one introduced in 
Ref.~\cite{toth1}: We derive bounds for the internal
energy $U=\mean{H}$ which have to hold for states without
multipartite entanglement, violation of these bounds implies the
presence of multipartite entanglement in the system. We investigate 
{in detail} two one-dimensional spin models: the 
Heisenberg model and the isotropic XY model.
Our results lead to the insight that for these models
even at modest temperatures the correlations cannot be 
explained without assuming the presence of genuine multipartite 
entanglement. However, we would like to stress that in our approach
we do not assume that the spin system is in the ground state or 
in a thermal state, our theorems can also be applied to states
out of equilibrium.

Our paper is organized as follows: In Section II we introduce the 
basic definitions of multipartite entanglement we want to apply to 
spin systems. We first recall the notion $n$-separability and 
genuine multipartite entanglement. Then, we also introduce the 
notion of $k$-producibility, which is well suited for the investigation
of multipartite entanglement in macroscopic systems. 
In Section III we apply these terms to the anti-ferromagnetic 
Heisenberg chain. We calculate energy bounds below which either 
reduced states are genuine multipartite entangled or the total state
requires multipartite entanglement for its creation.
This shows that already at temperatures of $kT \approx J$
the effect of multipartite entanglement cannot be neglected.
In Section IV we demonstrate that our proofs also work 
for other models by calculation  similar thresholds for the 
XY model. Finally, we summarize our results and name some open 
problems.

\section{Notions of multipartite entanglement}

Let us start by clarifying the terms we will use to
classify multipartite entanglement in spin chains.
For a pure state $\ket{\psi}$ of a quantum systems
of $N$ parties we may ask for a given $n\leq N$,
whether it is possible to cluster the parties into $n$ groups,
such that $\ket{\psi}$ is a
product state with respect to this partition,
\be
\ket{\psi}=\ket{\phi_1} \otimes\ket{\phi_2}\otimes ...\otimes \ket{\phi_n}.
\ee
If this is the case, we call the state $\ket{\psi}$
{\it $n$-separable.} A state which is $N$-separable is a
product state with respect to all subsystems, these we
call {\it fully separable.} If a state is not $2$-separable,
we call it {\it genuine $N$-partite entangled.}
For a mixed state described by a density matrix $\vr$ these
terms can be extended through convex combination.
If we can write $\vr=\sum_i p_i \ketbra{\psi_i}$ with
$p_i \geq 0, \sum_i p_i =1$ and $n$-separable $ \ket{\psi_i}$
we call $\vr$ $n$-separable. Physically, this means that a
production of $\vr$ requires only $n$-separable pure states
and mixing \cite{bem1}.

Besides asking for $n$-separability, we may also ask
questions like ``Do two party entangled states suffice to
create the  mixed state $\vr$?'' This leads to the
following definition. We call a state $\ket{\psi}$
{\it producible by $k$-party entanglement}
(or $k$-producible, for short) if we can write
\be
\ket{\psi}=\ket{\phi_1}\otimes\ket{\phi_2}\otimes
...\otimes \ket{\phi_m}
\ee
where the $\ket{\phi_i}$ are states of maximally $k$
parties. So, in this definition $m \geq N/k$ has to
hold. This definition means that it suffices to
generate specific $k$-party entanglement to arrive at the
state $\ket{\psi}.$ Conversely, we call a state
{\it containing genuine $k$-party entanglement}
if it is not producible by $(k-1)$-party entanglement.
This definition can be extended to mixed states as before
via convex combinations. Again, a mixed state which is
$k$-producible requires only the generation of $k$-party
entangled states and mixing for its production.
Consequently, {a mixed state $\vr$ contains $k$-party
entanglement, iff the correlations cannot be explained
by assuming the presence of $(k-1)$-party entanglement
only.}

Obviously, there are some relations between the notions
of $k$-producibility and $n$-separability. For instance,
the states containing $N$-party entanglement are just the
genuine multipartite entangled states and the $1$-producible
states are the fully separable states.
Furthermore, a state of  which some reduced state of $m$
parties is genuine $m$-partite entangled, contains
$m$-partite entanglement, while the converse is in general
not true \cite{remark1}.

In the thermodynamic limit $N \gg 1,$ however, the notion
of $n$-separability for the total system becomes problematic.
This is mainly due to two reasons. First, it does not take
into account how many particles are entangled. A biseparable
state can be a product state for two partitions of equal size
or just one qubit might be separated from a large genuine
multipartite entangled state; these cases are not distinguished. 
Second, statements about $n$-separability require the 
exact knowledge of $N,$ --- a precondition which is usually 
not fulfilled in realistic situations. 
In stark contrast, the notion of
$k$-producibility is designed to be sensitive to
the question ``How many qubits are entangled?''
Also, deciding whether a state is $k$-producible 
does not require the exact
knowledge of $N.$ If one controls only $k \ll N$
subsystems, one may still conclude that the
total state contains $k$-party entanglement,
e.g. if the reduced state is genuine $k$-party entangled.

\section{The Heisenberg model}
Let us start the discussion of spin models with the anti-ferromagnetic
isotropic Heisenberg model. Here we assume a one-dimensional chain of
$N$ spin $1/2$ particles with periodic boundary conditions. The
Hamiltonian of this model is given by
\be
H_H=
J \sum_{i=1}^N
\sigma_x^{(i)}\sigma_x^{(i+1)}
+\sigma_y^{(i)}\sigma_y^{(i+1)}
+\sigma_z^{(i)}\sigma_z^{(i+1)}
\label{XXXmodel}
\ee
where $\sigma_x^{(i)}$ denotes the Pauli matrix $\sigma_x,$
acting on the $i$-th qubit and $J>0$ the coupling between the
spins. For simplicity, we assume that $N$ is
even. For completeness, we always mention also known
results for fully separable states \cite{toth1, multi1}.
We have the following theorems:

{\bf Theorem 1.}
Let  $\vr$  be an $N$ qubit state of a system described
by the Heisenberg Hamiltonian in Eq.~(\ref{XXXmodel}).
If
\be
{\mean{H_H}} < - JN =: C_{R2}
\label{XXXmodel1c}
\ee
there exists two neighboring qubits $i, i+1$
in the chain such that the reduced state
$\vr_{i,i+1}$ is a two-qubit entangled state.
Furthermore, if
\be
{\mean{H_H}} < - \frac{1+\sqrt{5}}{2}JN \approx - 1.618 J N =: C_{R3}
\label{XXXmodel1b}
\ee
then there exist three neighboring qubits $i, i+1, i+2$
such that the reduced state $\vr_{i,i+1,i+2}$
of these qubits is genuine tripartite entangled.

{\it Proof.} It suffices to prove these bounds
for pure states. To prove first Eq.~(\ref{XXXmodel1b})
we write $H_H/J=W_{123}+W_{345}+W_{567}+...$ with
$W_{ijk}
=
\sigma_x^{(i)}\sigma_x^{(j)}
+\sigma_y^{(i)}\sigma_y^{(j)}
+\sigma_z^{(i)}\sigma_z^{(j)}
+\sigma_x^{(j)}\sigma_x^{(k)}
+\sigma_y^{(j)}\sigma_y^{(k)}
+\sigma_z^{(j)}\sigma_z^{(k)}.
$
The idea is to view the $W_{ijk}$ as entanglement witnesses,
detecting genuine tripartite entanglement on the qubits $i,j,k.$
Let us denote $x_i=\mean{\sigma_x^{(i)}}, x_i x_j =
\mean{\sigma_x^{(i)}\sigma_x^{(j)}}$ etc. and consider a
biseparable state $\ket{\psi}=\ket{\phi_i}\ket{\phi_{jk}}.$
We have
$|\mean{W_{ijk}}| =
| x_i \cdot x_j
+ y_i \cdot y_j
+ z_i \cdot z_j
+
x_jx_k
+ y_j y_k
+ z_j z_k |
\leq
\sqrt{
x_j^2 + y_j^2
+ z_j^2 }
+
|
x_jx_k
+ y_j y_k
+ z_j z_k
|.
$
For the two-qubit state $\ket{\phi_{jk}}$ the
first term is the purity of the reduced state
and invariant under local unitaries, while the
second is maximal, when the $3 \times 3$ correlation
matrix $\lambda_{\mu \nu} = \mu_j \nu_k$ for
$\mu, \nu=x,y,z$ is diagonal. Expressing a general
state in this form (see Eq.~(14) in Ref.~\cite{oviedo})
leads to $|\mean{W_{ijk}}|\leq 1+\sqrt{5}.$ Since there
are $N/2$ of the $W_{ijk},$ Eq.~(\ref{XXXmodel1b}) implies
that one of the reduced three-qubit states cannot be
biseparable. Finally, Eq.~(\ref{XXXmodel1c}) can be proved
similarly, using
$|x_i \cdot x_j + y_i \cdot y_j + z_i \cdot z_j| \leq 1$
for separable two-qubit states \cite{toth1, multi1}.
$\qed$

{\bf Theorem 2.} Let  $\vr$  be an $N$ qubit state
as in Theorem 1. If $\vr$ is $1$-producible, then
\be
{\mean{H_H}} \geq  -J N =: C_{C2}
\label{XXXmodel2}
\ee
holds, while for 2-producible states
\be
{\mean{H_H}} \geq  - \frac{3}{2}J N =: C_{C3}
\label{XXXmodel3}
\ee
holds. Thus, if ${\mean{H_H}} < C_{C3}$ the state contains
genuine tripartite entanglement.

{\it Proof.}
The bound Eq.~(\ref{XXXmodel2}) follows from
Eq.~(\ref{XXXmodel1c}) and has already been derived in
\cite{toth1, multi1}. Let us first consider a two-producible 
pure state, where neighboring spins
are allowed to be entangled. I.e., we have a state of the type
$\ket{\psi} = \ket{\phi_{12}}\ket{\phi_{34}}
... \ket{\phi_{N-1,N}},$ where the state $\ket{\phi_{12}}$ is a
state of the qubits 1 and 2, etc. Let us define for the state
$\ket{\psi}$ two vectors with $6N$ real components each, via
\begin{eqnarray}
\vec{v}_1 &:=&
([1],[1\!\!:\!\!2],[2],[\eins],[5],[5\!\!:\!\!6],[6], ...,[\eins]),
\nonumber
\\
\vec{v}_2 &:=&
([N],[\eins],[3],[3\!\!:\!\!4],[4],[\eins],[7], ...,[N-1\!\!:\!\!N]),
\nonumber
\end{eqnarray}
where the symbols $[...]$ stand always for three entries, namely
$[i]=x_i,y_i,z_i,$ and $[i\!\!:\!\!j]=x_i x_j, y_i y_j, z_i z_j$
and $[\eins]=1,1,1.$ Since the $\vec{v}_j$ have $6N$ entries, in
$\vec{v}_1$ (and similarly for $\vec{v}_2$) some coefficients
(like $x_1$) may appear twice, namely iff $N\in4 \mathbbm{Z}.$
It is straightforward to see that for the given
state $2 \cdot \mean{H_H}= J \vec{v}_1\cdot \vec{v}_2$  holds.
Now, we need
$\chi = x_i^2+y_i^2+ z_i^2+ (x_i x_{i+1})^2+ (y_i y_{i+1})^2+ (z_i
z_{i+1})^2+ x_{i+1}^2+ y_{i+1}^2+ z_{i+1}^2 \leq \chi_{max}=3,$
which holds for any two-qubit state $\vr,$ due to $(1+\chi)/4 \leq
Tr(\vr^2)\leq  1.$ Based on that, we have
$\Vert \vec{v}_1 \Vert^2 \leq
(N/2)(\chi_{max}+3)=3N.$ The same bound holds for $\Vert
\vec{v}_2\Vert^2.$ So, due to the Cauchy-Schwarz inequality
we have $|\mean{H_H}| \leq 1/2 J \Vert \vec{v}_1\Vert\cdot
\Vert \vec{v}_2\Vert \leq 3/2 J N$ which proves the claim.

Now we have to consider arbitrary two-producible states.
A general two-producible pure state is always a tensor product
of two-qubit states $\ket{\phi_{ij}}$ of the qubits $i$ and $j$
and single qubit states $\ket{\phi_{k}}$. If for one of the
two-qubit states $\ket{\phi_{ij}}$ the qubits $i$ and $j$ are
not neighboring, we can replace it by two one-qubit reduced states
$\vr_i \otimes \vr_j,$ since in this case, the Hamiltonian is
only sensitive to the reduced states.
Furthermore, if two neighboring single qubit states
$\ket{\phi_{i}} \otimes \ket{\phi_{j}},$ appear,
we replace them by $\ket{\phi_{ij}},$ since $\ket{\phi_{ij}}$
is allowed to be separable.

Thus it suffices to prove the bound for a state where entanglement
is only present between neighboring qubits and the single qubit
states are isolated in the sense that if $\ket{\phi_{k}}$
appears, then $\ket{\phi_{k-1}}$ and $\ket{\phi_{k+1}}$ do not
appear, e.g. $\ket{\psi} = \ket{\phi_{12}}\ket{\phi_{3}}
\ket{\phi_{45}}\ket{\phi_{67}}\ket{\phi_{8}}.$
Let $M$ be the (even) number of isolated qubits.
We can associate to any isolated qubit one neighboring two-qubit
state (on the left or right) to form a three-party group. There is
an ambiguity in doing that, and we can choose the three-qubit
groups in such a way that the number of two-qubit groups between
the three-qubit groups is always even. Then, the state can be
viewed as a sequence $g_1,g_2,g_3,g_4...$ of $M$ three-party
groups and $(N-3M)/2$ two party groups, all in all there are
$(N-M)/2$ groups. We can double this sequence by setting
$g_{(N-M)/2+k}=g_k$ etc. In the example this grouping can be
$(12),(345),(678),(12),(345),(678).$ Now we define the vectors
$\vec{v}_1$ and $\vec{v}_2$ as follows: $\vec{v}_1$ collects
terms from $g_1,g_3,g_5...$ while $\vec{v}_2$ consists of
terms from $g_2,g_4,...$ In detail, we have
\bea
\vec{v}_1
&=&
([g_1],[g_1|g_2],[\eins],[g_2|g_3],[g_3],[g_3|g_4],[\eins],
...), \nonumber
\\
\vec{v}_2
&=&
([\eins],[g_1|g_2],[g_2],[g_2|g_3],[\eins],[g_3|g_4],[g_4],
...), \nonumber
\eea
where $[g_1]$ denotes the three terms $x_i
x_j,y_i y_j, z_i z_j$ if $g_1$ is a two-qubit group and
corresponds to the four terms $x_i x_j, y_i y_j, z_i z_j, (x_j x_k
+ y_j y_k + z_j z_k) $ when $g_1$ is a three-qubit group with the
isolated qubit $k.$ $[g_1|g_{2}]$ etc. denotes the coupling terms
between the groupings $g_1$ and $g_2,$ in the Hamiltonian, e.g.
$x_i,y_i,z_i$ for $v_1$ and $x_{i+1},y_{i+1}, z_{i+1}$ for $v_2$
or vice versa.
The symbol $[\eins]$ denotes a sequence of three or four times
``1'', depending on whether in the other vector there is a two- or
a three-qubit group.

Again, we have $2 \cdot \mean{H}= J \vec{v}_1 \cdot \vec{v}_2$
and it remains to bound $\Vert \vec{v}_i \Vert^2.$ First, note
that by construction $\vec{v}_1$ and $\vec{v}_2$ contain the same
number of two and three-qubit groups, namely
$(N-3M)/2$ two-qubit groups and $M$ three-qubit groups
each. Also, both contain $3 (N-3M)/2 +4 M$ times the ``1''.
Then, note that for states of the type
$\ket{\psi} = \ket{\phi_{ij}} \ket{\phi_{k}}$
the bound
$x_i^2+ y_i^2+ z_i^2+ (x_i x_j)^2+ (y_i y_j)^2 + (z_i z_j)^2 +
(x_j \cdot x_k + y_j \cdot y_k + z_j \cdot z_k)^2+ x_k^2+ y_k^2+ z_k^2 \leq 5$
is valid. So we have
$\Vert \vec{v}_i \Vert^2 \leq (4+5) M + (3+3)(N-3M)/2 =3N, $
which proves the claim.
$\qed$

To start the discussion, first note that the bounds
in Eqs.~(\ref{XXXmodel2}, \ref{XXXmodel3}) are sharp.
Equation (\ref{XXXmodel3}) is saturated for the singlet chain
$\ket{\psi} = \ket{\psi^-}\ket{\psi^-}...\ket{\psi^-}$
where $\ket{\psi^-}=(\ket{01}-\ket{10})/\sqrt{2}.$

In general, any energy bound $\mean{H_H} \geq C_X$
corresponds to a certain temperature $T_X.$
Below this temperature the state has with certainty
some degree of entanglement. Corresponding to the
Eqs.~(\ref{XXXmodel1c}, \ref{XXXmodel1b}, \ref{XXXmodel2},
\ref{XXXmodel3}) there are thus the temperatures $T_{R2},
T_{R3}, T_{C2}$ and $T_{C3}$ below which either reduced
states of two or three parties are entangled or the
total state contains two- or three-party entanglement.
Obviously, $T_{R2}= T_{C2} > T_{C3} > T_{R3}$
has to hold here.

Let us estimate these temperatures. For small $N$ one can solve 
the Heisenberg model by diagonalizing
$H_H$ numerically \cite{bethe}. 
Then, the threshold temperatures can directly
be computed. Results are shown in Fig.~1.
\begin{figure}[t]
\centerline{\epsfxsize=0.9\columnwidth
\epsffile{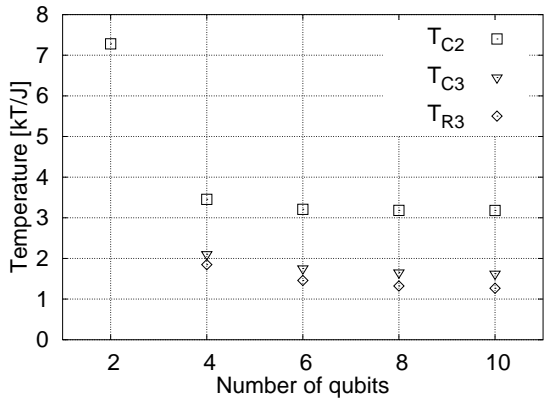} }
\caption{ Threshold temperatures $T_{R3}, T_{C2}$ and $T_{C3}$ for
small Heisenberg chains up to ten qubits.
See text for details.
}
\label{fig_settings}
\end{figure}
As expected, the values for $ T_{C2}=T_{R2}$ coincide with the
ones of Ref.~\cite{multi1}. The given
values for $T_{C3}$ and
$T_{R3}$ show that in the Heisenberg chain of ten spins at
$kT \approx J$ multipartite entanglement plays a
role, namely at least one reduced state is genuine tripartite
entangled and the total state contains tripartite entanglement.

In the thermodynamic limit $N \gg 1$ the ground state energy
of the Heisenberg model is known to be $E_0/N= - J (4 \ln 2 -1)
\approx - 1.773 J$ \cite{hulthen}.
Thus three qubits can be found  such that their reduced state 
is genuine tripartite entangled. One can infer from numerical 
calculations \cite{xiang, bethe}
that the threshold temperatures are  determined by
$k T_{C3}\approx 1.61 J$ and $k T_{R3} \approx 1.23 J$
in agreement with the values of the ten-qubit Heisenberg 
chain.

Finally, note that our results also shed light on the characterization
of the ground state itself. For instance, they show that
the ground state cannot be a GHZ state since for that state the
reduced three-qubit states are separable. Finally, note that
when a state is translationally  invariant, Eq.~(\ref{XXXmodel1b})
guarantees that {\it all} reduced three-qubit states are genuine
tripartite entangled.

\section{The XY model}
Let us now investigate the isotropic XY model. 
The Hamiltonian of this
model is given by
\be
H_{XY}=
J \sum_i
\sigma_x^{(i)}\sigma_x^{(i+1)}
+\sigma_y^{(i)}\sigma_y^{(i+1)}.
\label{XXmodel}
\ee
Again, we assume periodic boundary conditions and an even number
of spins. For this model we have:

{\bf Theorem 3.}
Let  $\vr$  be an $N$ qubit state whose dynamics
is governed by the Hamiltonian in Eq.~(\ref{XXmodel}).
If $\vr$ is one-producible, then
\be
{\mean{H_{XY}}} \geq  -J N
\label{XXmodel1}
\ee
holds. If ${\mean{H_{XY}}} < -J N$ this implies that
there are two neighboring qubits such that their
reduced state is entangled.
For two-producible states
\be
{\mean{H_{XY}}} \geq  - \frac{9}{8}J N
\label{XXmodel2}
\ee
holds. If ${\mean{H_{XY}}} < - {9}/{8}J N$ the state contains thus
tripartite entanglement and if
\be
{\mean{H_{XY}}} < - \frac{1+\sqrt{2}}{2} J N  \approx - 1.207 J N
\label{XXmodel3}
\ee
then there are three neighboring qubits such that their
reduced state is genuine tripartite entangled.

{\it Proof.}
The proofs for the XY model are similar to the ones for the 
Heisenberg model, so we can make it short. Eq.~(\ref{XXmodel1}) 
can be proved in the same manner as for the Heisenberg model 
\cite{toth1}. Now, let us first show the bound (\ref{XXmodel2}) 
for a state of the type
$\ket{\psi} = \ket{\phi_{12}}\ket{\phi_{34}}... \ket{\phi_{N-1,N}}.$
Again, we define two vectors $\vec{v}_1$ and $\vec{v}_2,$
with now $4N$ entries via
$ \vec{v}_1
:=
([1],[1\!\!:\!\!2],[2],[\eins],[5],
[5\!\!:\!\!6],[6],[\eins], ...,[\eins])$
and
$\vec{v}_2 :=
([N],[\eins],[3],[3\!\!:\!\!4],[4],[\eins],
[7],[7\!\!:\!\!8], ...,[N-1\!\!:\!\!N]),$
where now $[i]=x_i,y_i$ and $[i\!\!:\!\!j]=
\alpha \beta_{1,2},\alpha \beta_{1,2}$ and
$[\eins]= \alpha^{-1},\alpha^{-1}$ with
$\alpha=\sqrt{4/3}$ and $\beta_{i,j} =(x_i x_j+y_i
y_j)/2.$
Again we have $2\mean{H_{XY}}=J\vec{v}_1\cdot \vec{v}_2$
and due to the fact that
$x_i^2+y_i^2+2 \alpha^2 \beta_{i,j}^2 + y_j^2 +
x_j^2 \leq \xi_{max}=3$
we have that $\Vert \vec{v}_1\Vert^2 \leq N/2
(\xi_{max}+2/\alpha^2)=9/4N,$ which proves the claim.
General two-producible states can be treated as shown
in the proof of Theorem 2, now for states
of the type $\ket{\psi_{ij}}\otimes\ket{\psi_k}$ the
bound $x_i^2+y_i^2+2\alpha^2 \beta_{i,j}^2 +
\alpha^2(x_j \cdot x_k+y_j \cdot y_k)^2 +y_k^2 + x_k^2
\leq x_i^2+y_i^2+2 \alpha^2 \beta_{i,j}^2 +\alpha^2(x_j^2+y_j^2) +
1 \leq 9/2$ must be used.
Finally, Eq.~(\ref{XXmodel1}) can be
proved as Eq.~(\ref{XXXmodel1b}), using $
\sqrt{x_j^2 + y_j^2} + |x_j x_k + y_j y_k| \leq 1+\sqrt{2}.$
$\qed$

First note that again the bounds in
Eqs.~(\ref{XXmodel1},\ref{XXmodel2}) are sharp. In
Eq.~(\ref{XXmodel2}) equality holds for the state
$\ket{\psi}=\ket{\phi_{12}}\ket{\phi_{34}}....\ket{\phi_{N-1,N}}$
where $\ketbra{\phi_{i,i+1}} = 1/4 \cdot (\eins\otimes \eins -
\sigma_x  \otimes \sigma_x - 1/2 \cdot [\sigma_y  \otimes \sigma_y
+ \sigma_z  \otimes \sigma_z] - \sqrt{3}/2 \cdot[\eins \otimes
\sigma_x - \sigma_x \otimes \eins]) $ for all odd $i.$ This
corresponds to a chain of non-maximally entangled two-qubit states.
This also motivates the definition of $\alpha$ and $\beta_{i,j}$
in the proof of Eq.~(\ref{XXmodel2}). $\alpha$ and $\beta_{i,j}$
were chosen such that for the state $\ket{\psi}$ we have
$\vec{v}_1=\vec{v}_2,$ thus the Cauchy-Schwarz inequality 
is sharp.

In the thermodynamic limit the XY model is exactly solvable
\cite{katsura}. The ground state energy is
$E_0/N= - 4J / \pi \approx -1.273 J.$ Thus the ground state contains
tripartite entanglement and fulfills the condition of
Eq.~(\ref{XXmodel3}). Furthermore, the results of Ref.~\cite{katsura}
imply that if $kT < 0.977 J$ the mixed state contains
tripartite entanglement and if $kT < 0.668 J$  some reduced state
is genuine tripartite entangled.

\section{Conclusion}

In conclusion, we showed for two important spin models
that the internal energy can be used as a signature for
the presence of multipartite entanglement in these models.
Based on this, we computed threshold temperatures below
which any realistic description of the system cannot neglect 
the effects of multipartite entanglement. Our results may 
stimulate the research on entanglement in phase 
transitions, since they suggests the use of multipartite 
entanglement measures as a tool for the investigation 
of phase transitions in these regimes \cite{italia}.


A natural continuation of the present work is the extension of 
the Theorems presented here to other spin systems, e.g. higher 
dimensional or frustrated systems. Furthermore, it is very 
tempting to analyze the results of performed experiments whether
they can be interpreted as giving evidence for multipartite
entanglement, similarly as it has been done for bipartite
entanglement in Ref.~\cite{experiments}. Here, we leave this 
as an open problem.

We thank J.I.~Cirac, W.~Dür, J.J.~Garc\'{\i}a-Ripoll,
A.~Miyake, M.~Plenio and especially T.~Xiang for discussions.
This work has been supported by the EU (MEIF-CT-2003-500183,
IST-2001-38877, -39227) and the DFG.


\begin{thebibliography}{99}

\bibitem{oster}
T. J. Osborne and M. A. Nielsen,
Phys. Rev. A {\bf 66}, 032110 (2002).
A. Osterloh, L. Amico, G. Falci, and R. Fazio,
Nature {\bf 416}, 608 (2002).

\bibitem{localizable}
F. Verstraete, M. Popp and  J.I. Cirac, 
Phys. Rev. Lett. {\bf 92}, 027901 (2004); 
B.-Q. Jin and V.E. Korepin, 
Phys. Rev. A {\bf 69}, 062314 (2004).
M. Popp, F. Verstraete, M.A. Martin-Delgado,  and J.I. Cirac
Phys. Rev. A {\bf 71}, 042306 (2005).


\bibitem{arealaws}
G. Vidal, J.I. Latorre, E. Rico, and A. Kitaev,
Phys. Rev. Lett. {\bf 90}, 0227902 (2003);
V.E. Korepin, 
Phys. Rev. Lett. {\bf 92}, 096402 (2004);
B.-Q. Jin and V.E. Korepin,  
J. Stat. Phys. {\bf 116}, 79 (2004);
A.R. Its, B.-Q. Jin and V.E. Korepin,  
J. Phys. A: Math. Gen. {\bf 38}, 2975 (2005);
M.B. Plenio, J. Eisert, J. Dreißig, and M. Cramer,
Phys.Rev.Lett. {\bf 94} 060503 (2005).

 

\bibitem{vidal}
G. Vidal, Phys. Rev. Lett. {\bf 93}, 040502 (2004);
F. Verstraete, D. Porras and  J.I. Cirac, Phys. Rev. Lett. 
{\bf 93}, 227205 (2004); 
U. Schollwöck, Rev. Mod. Phys. {\bf 77}, 259 (2005).

\bibitem{toth1}
G. T\'oth, Phys. Rev. A {\bf 71}, 010301(R) (2005);
\v{C}.~Brukner and V.~Vedral, quant-ph/0406040;
M.R.~Dowling, A.C.~Doherty, and S.D.~Bartlett 
Phys. Rev. A {\bf 70}, 062113 (2004); L.-A. Wu, S.~Bandyopadhyay,
M.S. Sarandy, and D.A. Lidar,
Phys. Rev. A {\bf 72}, 032309 (2005)

\bibitem{buttiker}
For other works relating entanglement to macroscopic 
observables, see, e.g.
S. Ghosh, T.F. Rosenbaum, G. Aeppli and S.N. Coppersmith,
Nature (London) {\bf 425}, 48 (2003); 
M. Wiesniak, V. Vedral and \v{C}.~Brukner,
quant-ph/0503037;
for an approach relating fluctuations of subsystem 
energies to entanglement see 
A.N. Jordan and M. B\"uttiker
Phys. Rev. Lett. {\bf 92}, 247901 (2004).

\bibitem{remark5} In these works, the conclusion that the
state is entangled relies on the assumption that the model
for the physical system is valid. Strictly speaking, one
proves the existence of entanglement in a certain model of
statistical mechanics.

\bibitem{multi1}
X. Wang, Phys. Rev. A {\bf 66}, 044305 (2002).

\bibitem{multi2}
L.F. Santos, 
Phys. Rev. A {\bf 67}, 062306 (2003);
P.~\v{S}telmachovi\v{c} and V. Bu\v{z}ek, 
Phys. Rev. A {\bf 70}, 032313 (2004);
T.-C. Wei, D. Das, S. Mukhopadyay, S. Vishveshwara,  and P.M. Goldbart, 
Phys. Rev. A {\bf 71}, 060305(R) (2005)
D. Bru{\ss}, N. Datta, A. Ekert, L.C. Kwek, and C. Macchiavello, 
Phys. Rev. A {\bf 72}, 014301 (2005);
C.~Lunkes, \v{C}.~Brukner and V.~Vedral,
Phys. Rev. Lett. {\bf 95}, 030503 (2005).
  
 

\bibitem{duer}
W. Dür, G. Vidal, and J.I. Cirac,
Phys. Rev. A {\bf 62}, 062314 (2000);
A. Miyake
Phys. Rev. A  {\bf 67}, 012108 (2003);
F.~Verstraete, J. Dehaene, and B. De Moor,
Phys. Rev. A {\bf 68}, 012103 (2003).


\bibitem{acin}
A. Ac\'{\i}n, D. Bru{\ss}, M. Lewenstein, and A. Sanpera,
Phys. Rev. Lett. {\bf 87}, 040401 (2001).

\bibitem{bem1}
Note that a mixed state
$\vr=\sum_i p_i \ketbra{\psi_i}$ can usually
be written as a convex combination in many ways.
Thus a state which was generated by mixing entangled states
might nevertheless be separable, since on may create the same
state also by mixing separable states. Note also that for an
$n$-separable state the states $\ket{\psi_i}$ may be
$n$-separable with respect to different partitions.

\bibitem{remark1}
This follows from the fact that there are
three-qubit states which are separable for each
fixed partition but not fully separable \cite{acin}.
These states contain two-party entanglement,
however, any reduced two-qubit state is separable.

\bibitem{oviedo}
O. Gühne, P. Hyllus, D. Bru{\ss}, A. Ekert, M. Lewenstein, 
C. Macchiavello and A. Sanpera, J. Mod. Opt. {\bf 50}, 1079 (2003).

\bibitem{bethe} 
Alternatively, one may also use the Bethe ansatz to derive the 
thermodynamic properties. See 
H.A. Bethe, 
Z. Physik {\bf 71}, 205 (1931);
C.N. Yang and C.P. Yang,
J. Math. Phys. {\bf 10}, 1115 (1969);
M. Takahashi, 
Prog. Theor. Phys. {\bf 45}, 401 (1971)
and the book by
M. Takahashi,
{\it Thermodynamics of one-dimensional solvable models}
(Cambridge University Press, 1999).

\bibitem{hulthen}
L. Hulth\'en,
Arkiv Math. Astron. Fys. {\bf 26A} 1 (1938).

\bibitem{xiang}
T. Xiang, private communication and
Phys. Rev. B {\bf 58}, 9142 (1998).

\bibitem{katsura}
S. Katsura,
Phys. Rev. {\bf 127}, 1508 (1962).

\bibitem{italia} For research in this direction see, e.g.
T. Roscilde, P. Verrucchi, A. Fubini, S. Haas, and V. Tognetti,
quant-ph/0412098.

\bibitem{experiments} See,
\v{C}.~Brukner, V. Vedral,  and A.~Zeilinger,
quant-ph/0410138. Note that the system
investigated there can be described well by a
two-producible state, thus it is not suitable
for the search for multipartite entanglement.



\end{thebibliography}
\end{document}